\documentclass[runningheads]{llncs}
\pdfoutput=1
\usepackage{graphicx}
\usepackage[ansinew]{inputenc}
\usepackage[T1]{fontenc}
\usepackage{amsfonts}         
\usepackage{amsmath}
\usepackage{amssymb}
\usepackage{color}
\usepackage{lmodern}
\usepackage{array}
\usepackage{graphicx}
\usepackage{fancyvrb}
\usepackage{multicol}
\usepackage{multirow}
\usepackage{float}
\usepackage{bm}
\usepackage{bbm}

\usepackage{algorithm}
\usepackage[noend]{algpseudocode}
\usepackage{varwidth}

\floatname{algorithm}{Algorithm}

\algdef{SE}[SUBALG]{Indent}{EndIndent}{}{\algorithmicend\ }%
\algtext*{Indent}
\algtext*{EndIndent}

\makeatletter
\newcommand{\algmargin}{\the\ALG@thistlm}
\makeatother
\newlength{\whilewidth}
\algdef{SE}[parWHILE]{parWhile}{EndparWhile}[1]
  {\parbox[t]{\dimexpr\linewidth-\algmargin}{\hangindent\whilewidth\strut\algorithmicwhile\ #1\ \algorithmicdo\strut}}{\algorithmicend\ \algorithmicwhile}
\algnewcommand{\parState}[1]{\State
  \parbox[t]{\dimexpr\linewidth-\algmargin}{\strut #1\strut}}

\newcommand{\R}{\mathbb{R}}

\newcommand{\bigO}[1]{\mathcal{O}\left(#1\right)}

\begin{document}

\title{Efficient Autotuning of Hyperparameters in Approximate Nearest Neighbor Search}


\titlerunning{Efficient Autotuning of Hyperparameters in ANN Search}
%
\author{Elias J\"a\"asaari \inst{1} \and
Ville Hyv\"onen \inst{2}\inst{3} \and
Teemu Roos \inst{2}\inst{3}}
\authorrunning{E. J\"a\"asaari et al.}

\institute{Kvasir Ltd.
\email{elias.jaasaari@gmail.com} \and
University of Helsinki,  Department of Computer Science \and
Helsinki Institute for Information Technology (HIIT) \\
\email{ville.o.hyvonen@gmail.com}, \email{teemu.roos@cs.helsinki.fi}}
%
\maketitle              
\begin{abstract}
Approximate nearest neighbor algorithms are used to speed up nearest neighbor search in a wide array of applications. However, current indexing methods feature several hyperparameters that need to be tuned to reach an acceptable accuracy--speed trade-off. A grid search in the parameter space is often impractically slow due to a time-consuming index-building procedure. Therefore, we propose an algorithm for automatically tuning the hyperparameters of indexing methods based on randomized space-partitioning trees. In particular, we present results using randomized $k$-d trees, random projection trees and randomized PCA trees. The tuning algorithm adds minimal overhead to the index-building process but is able to find the optimal hyperparameters accurately. We demonstrate that the algorithm is significantly faster than existing approaches, and that the indexing methods used are competitive with the state-of-the-art methods in query time while being faster to build.


\end{abstract}

\section{Introduction}


\emph{Nearest neighbor search} is a common component of algorithms and pipelines in areas such as machine learning~\cite{Hassan2017,Wang2016}, computer vision~\cite{Bilen2015} and robotics~\cite{McBryde2018}. In modern applications the search is typically performed in high-dimensional spaces (100--10000 dimensions) over large data sets.

An exhaustive $k$-nearest neigbor ($k$-NN) search is often prohibitively slow in applications which either require real-time responses (see e.g.~\cite{Wang2016}) or run on a resource-constrained device (see e.g.~\cite{McBryde2018}). Hence, \emph{approximate} nearest neighbor (ANN) search is often used instead. ANN algorithms first build an index in an offline phase, after which the index can be used to perform $k$-NN queries in sublinear time in an online phase.
Most of the efficient algorithms fall into one of four categories: product quantization (PQ)~\cite{Jegou2011}, locality-sensitive hashing (LSH) \cite{Indyk1998,Dong2008}, graph-based methods~\cite{Malkov2016}, and tree-based methods~\cite{Silpa2008,Muja2014}. 

Because ANN algorithms are typically used as an auxiliary component of a pipeline, it can be important for a user that an algorithm requires minimal hand-tuning, especially if the type or size of the data can vary significantly. However, ANN algorithms typically have several hyperparameters which need to be tuned by a time-consuming grid search to achieve a given accuracy level or search time.

This problem is solved by an autotuning algorithm where the user specifies an accuracy level, and the tuning algorithm finds the optimal hyperparameter values. Previously, autotuning methods have been proposed for VP-trees \cite{Yianilos1993}, multi-probe LSH \cite{Dong2008}, $k$-means trees and RKD trees \cite{Muja2014}. In this paper, we propose an autotuning method that is significantly faster than these methods.



Our approach is based on exploiting the structure of randomized space-partitioning trees~\cite{Silpa2008,Muja2014,Dasgupta2015,Hyvonen2016}. ANN algorithms based on randomized space-partitioning trees have been used recently for example in machine translation~\cite{Hassan2017}, object detection~\cite{Bilen2015} and recommendation engines~\cite{Wang2016}.
 
Trees have several advantages: they are fast in high-dimensional spaces (see e.g. experiments in~\cite{Muja2014,Hyvonen2016}); they are simple to implement; they support easy insertion and deletion of points and they are independent, making the parallel implementation trivial. Also of great importance to us is that the structure of a tree-based index can be exploited to speed up the hyperparameter tuning.

Several types of randomized space-partitioning trees have been proposed for ANN search. Randomized $k$-d (RKD) trees~\cite{Silpa2008} with a priority queue search are used in the popular open-source library FLANN~\cite{Muja2014}. Random projection (RP) trees~\cite{Dasgupta2015} with a voting search have a stronger empirical performance than RKD trees with a priority queue search~\cite{Hyvonen2016}. However, a single principal component (PCA) tree has been found to be more accurate than a single RP tree~\cite{Verma2009}. The PCA tree has two problems: it is not randomized, and indexing is slow. To solve these problems, we design a randomized variant of the PCA tree.

Typically ANN algorithms are compared in terms of the accuracy--speed trade-off.
 However, for the algorithm to be useful in practice, the index building procedure must be efficient as well. We test three different types of trees (RKD, RP and randomized PCA) with two search methods (priority queue and voting) considering both the query stage and the index building stage. 

More specifically, in this article we:
\begin{list}{$\bullet$}{}
    \item Propose an autotuning algorithm to optimize the hyperparameters of tree-based ANN search, and demonstrate that it is faster and more accurate than existing autotuning methods for ANN algorithms.
    \item Compare experimentally the effect of a) the randomization strategy and b) the search method on the efficiency of randomized trees. In particular, we find RP trees combined with voting search to be the best-performing.
    \item Demonstrate that the best tree-based method is nearly on par with the state-of-the-art ANN algorithms when measured on the accuracy--speed trade-off, and faster when measured on the index building time.
\end{list}


\section{Approximate nearest neighbor search}

In \emph{$k$-nn search}, we have a data set  $\mathbf{x} = (x_1, \dots , x_n)$,
where each $x_i \in \mathcal{A}$, from which we want to find the indices $f(q)$ of the $k$ nearest neighbors for an arbitrary query point $q \in A$ measured by a dissimilarity measure $\mathrm{dis}(u, v):\mathcal{A}^2 \mapsto \R$.
 We assume the dissimilarity measure to be the Euclidean distance $
\left\|u - v\right\|_2$.


In \emph{approximate nearest neighbor} (ANN) search, it is sufficient that the $k$ points returned by the approximation algorithm are the true nearest neighbors of the query point only with high probability. We denote the returned points by $\hat{f}(q;\bm{\alpha}, \mathbf{r})$, where $\bm{\alpha}$ stands for the hyperparameters of the algorithm, and $\mathbf{r}$ stands for the realization of a set of random vectors used by the algorithm.

The accuracy of the approximation is measured by the \emph{error rate}
$
\mathrm{Err}(q;\bm{\alpha}, \mathbf{r}) = \frac{1}{k}\sum_{j=1}^k \mathbbm{1}(f_j(q) \notin \hat{f}(q;\bm{\alpha}, \mathbf{r})), 
\label{eq:error-rate}
$
which is the proportion of missed true nearest neighbors; the indices of the true nearest neighbors are denoted by $f(q) = (f_1(q), \dots, f_k(q))$. Equivalently, we can use \emph{recall}:  $ \mathrm{Rec}(q;\bm{\alpha}, \mathbf{r}) = 1 - \mathrm{Err}(q;\bm{\alpha}, \mathbf{r})$.

In addition to the error rate, we also  consider the query time, denoted $\mathrm{Time}(q;\bm{\alpha}, \mathbf{r})$, when assessing the performance of an ANN algorithm. The hyperparameter optimization problem can  be formulated in two ways: 

\begin{enumerate}
    \item 
Fix the expected error rate $e \in (0, 1)$ and find the hyperparameters $\bm{\alpha}$ that minimize $E\left[ \mathrm{Time}(Q; \bm{\alpha}, \mathbf{R}) \right]$ under the constraint $E\left[ \mathrm{Err}(Q;\bm{\alpha}, \mathbf{R})\right] \leq e$.
\item Fix the expected query time $t \in (0, \infty)$ and find the hyperparameters $\bm{\alpha}$ that minimize $E\left[ \mathrm{Err}(Q;\bm{\alpha}, \mathbf{R})\right]$ under the constraint $E\left[\mathrm{Time}(Q;\bm{\alpha}, \mathbf{R})\right] \leq t$.
\end{enumerate}

%
%

The expectations $E\left[\cdot\right]$ are over both the distribution of a query point $Q$ and the  random vectors $\mathbf{R}$. These expectations can be estimated using a validation set of query points and a generated sample of random vectors.




\section{Randomized space-partitioning trees}


\subsection{Index construction}
\label{sec:index-construction}
A binary space-partitioning tree recursively divides the data points into different cells with the assumption that nearby points fall into the same cells. At each branch of the recursion, the data set $\mathbf{x}$ is projected onto a chosen direction and assigned into one of the two child nodes by applying a split criterion. In practice we use the median split to ensure balanced trees. This process (Algorithm \ref{algo:grow-trees}) is continued at the child nodes until the maximum depth $\ell$ is met.

\begin{algorithm}[tb]
\begin{algorithmic}[1]
\raggedright
\Function{grow-tree}{depth, $\mathbf{x}, \ell, \bm{\psi}$}
  \If{depth == $\ell$}    
    \State \Return indices of points in $\mathbf{x}$ as a tree node
  \EndIf
  \State direction $\leftarrow$ \Call{generate-direction}{$\bm{\psi}$}
  \State p $\leftarrow$ \Call{project}{$\mathbf{x}$, direction}
  \State cut $\leftarrow$ \Call{split}{p}
   \State left $\leftarrow$ \Call{grow-tree}{depth + 1, $\mathbf{x}$[p $\leq$ cut], $\ell$}
  \State right $\leftarrow$ \Call{grow-tree}{depth + 1, $\mathbf{x}$[p > cut], $\ell$}
  \State \Return (left, right, cut, direction) as a tree node
\EndFunction
\end{algorithmic}
\caption{}
\label{algo:grow-trees}
\end{algorithm}

The type of a space-partitioning tree is determined by its choice of projection direction (see Figure 1 for an illustration on 2D data). In Algorithm \ref{algo:grow-trees}, each different type of tree implements its own version of the abstract function \textproc{generate-direction} which chooses this direction. Its argument $\bm{\psi}$ represents the tree-type dependent tuning parameters.



In randomized space-partitioning trees, the projection direction is chosen in a non-deterministic fashion. Randomized $k$-d (RKD) trees~\cite{Silpa2008} choose a coordinate direction uniformly at random from $m$ directions of the highest variance as the projection direction (we use $m = 5$ as suggested in~\cite{Silpa2008}). Another popular randomized variant is a random projection (RP) tree~\cite{Dasgupta2015} in which the projection direction is chosen uniformly at random from the $d$-dimensional unit sphere. We use a sparse version~\cite{Hyvonen2016}, in which only a proportion $a = 1 / \sqrt{d}$ of the components of the random vectors are non-zero.

If the first principal component of the data is used as the projection direction, the resulting data structure is a principal component (PCA) tree~\cite{Verma2009}. However, the original PCA trees are on the one hand deterministic which makes improving accuracy with multiple trees impossible, and on the other hand slow to compute, as computing exact PCA is costly. To speed up the computation, using gradient descent updates to approximate the first principal component of the data at each node of the tree has been suggested~\cite{Mccartin2012}. However, index construction still takes $\bigO{nd^2 (i + \ell)}$ time, where $i$ is the number of gradient descent updates. 




\label{sec:randomized-BST}
\begin{figure}[!h]
    \centering
    \includegraphics[scale=0.34,clip=yes,trim=50 20 50 30]{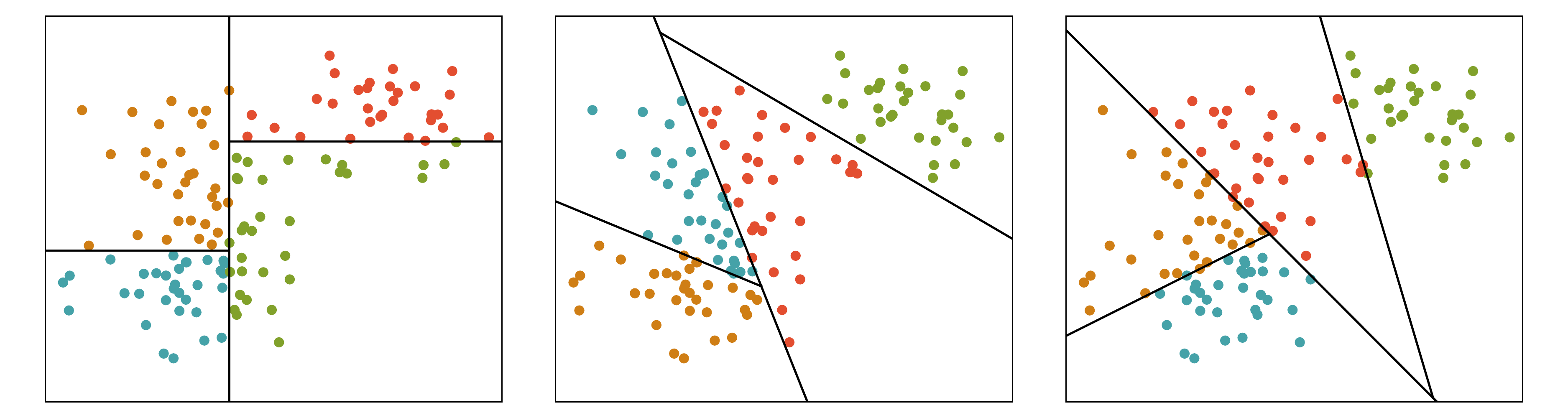}
    \caption{Different projection directions: $k$-d (left), RP (middle) and PCA (right).}
    \label{fig:my_label}
\end{figure}

We make PCA trees more practical for ANN search by modifying the gradient descent update\footnote{The gradient descent consistently converges with the learning rate $\gamma=0.01$ in all our experiments; we did not observe further tuning of the learning rate to be necessary.} to choose uniformly at random only $a = \sqrt{d}$ dimensions of the data at each node of the tree, and compute the estimated covariance matrix using only these dimensions. 
Growing a randomized PCA tree is an $\bigO{nd (i + \ell)}$ operation since now computing the sample covariance matrix takes only $\bigO{nd}$ operations. Considering only a sample of dimensions also ensures that the trees are randomized, allowing us to build multiple trees to increase accuracy.

\begin{figure}[!t]
    \centering
    \includegraphics[scale=0.42,clip=yes,trim=20 20 20 20]{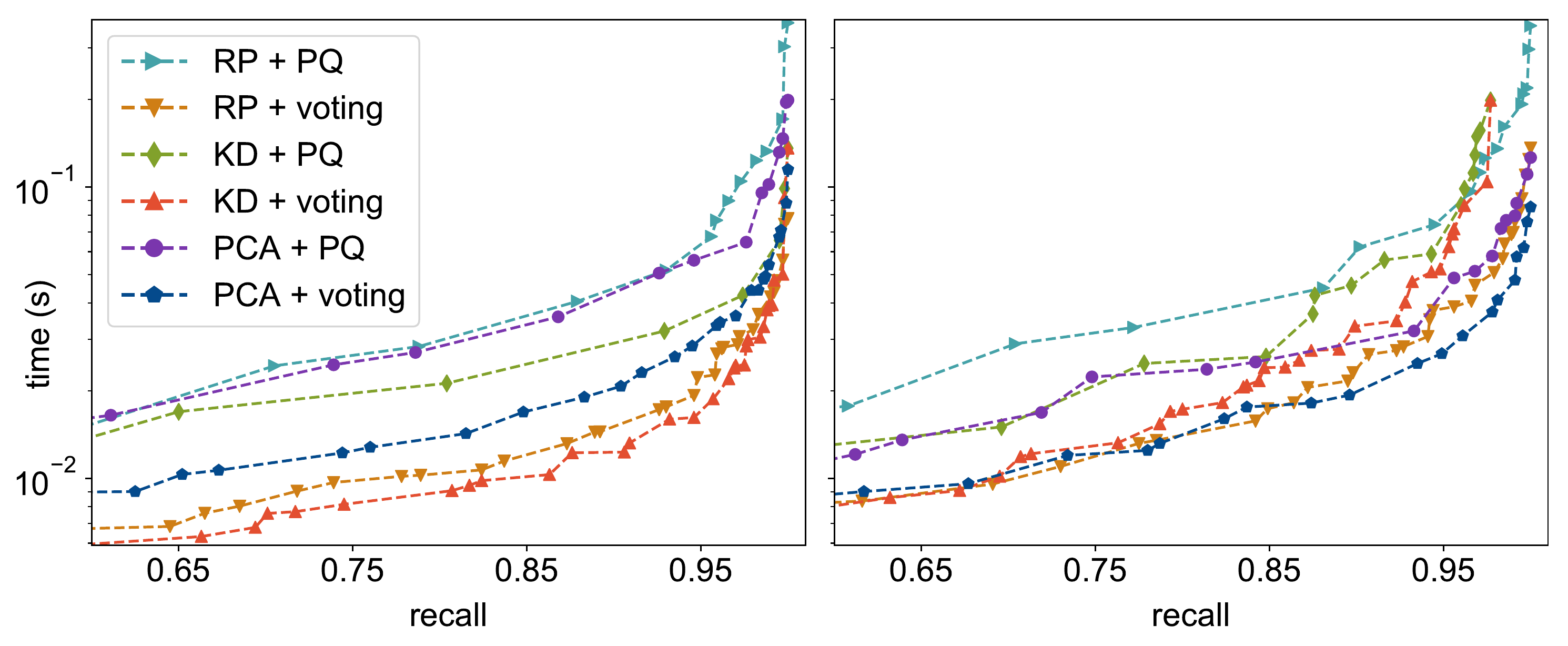}
    \caption{Recall vs. query time with different trees and search methods for MNIST (left) and Fashion-MNIST (right) for $k=10$. Towards bottom right is better.}
    \label{fig:pq_vs_voting}
\end{figure}

\subsection{ANN search using multiple trees}
\label{sec:tree-query}

To use an index consisting of $T$ randomized space-partitioning trees to find $k$ approximate nearest neighbors of a query point $q$, the query point is first routed down to a leaf at each of the trees: at each level the query point is first projected onto the saved projection direction and then routed into the left or the right child node depending on which side of the split point its projection falls. There are two strategies to choose the candidate set of points for which the true distances are evaluated: priority queue search and voting search. Both of these are independent of the randomization strategy used to grow the trees.



\subsubsection{Priority queue search}
In a priority queue search~\cite{Silpa2008}, a single priority queue, ordered according to the distance from the query point to the splitting hyperplanes, is maintained for all trees. When distances from the query point to all the points sharing a leaf with the query point are evaluated, $b$ extra branches are explored; the priority queue is used to choose the branches.

\subsubsection{Voting search}
In a voting search~\cite{Hyvonen2016}, distances are computed only to the subset of the points sharing a leaf with the query point. When a data point belongs to the same leaf as a query point in a tree, it gets a vote, and distances are evaluated only to the points that have at least $v$ votes. 


\begin{figure}[!t]
     \centering
     \includegraphics[scale=0.42,clip=yes,trim=20 20 20 20]{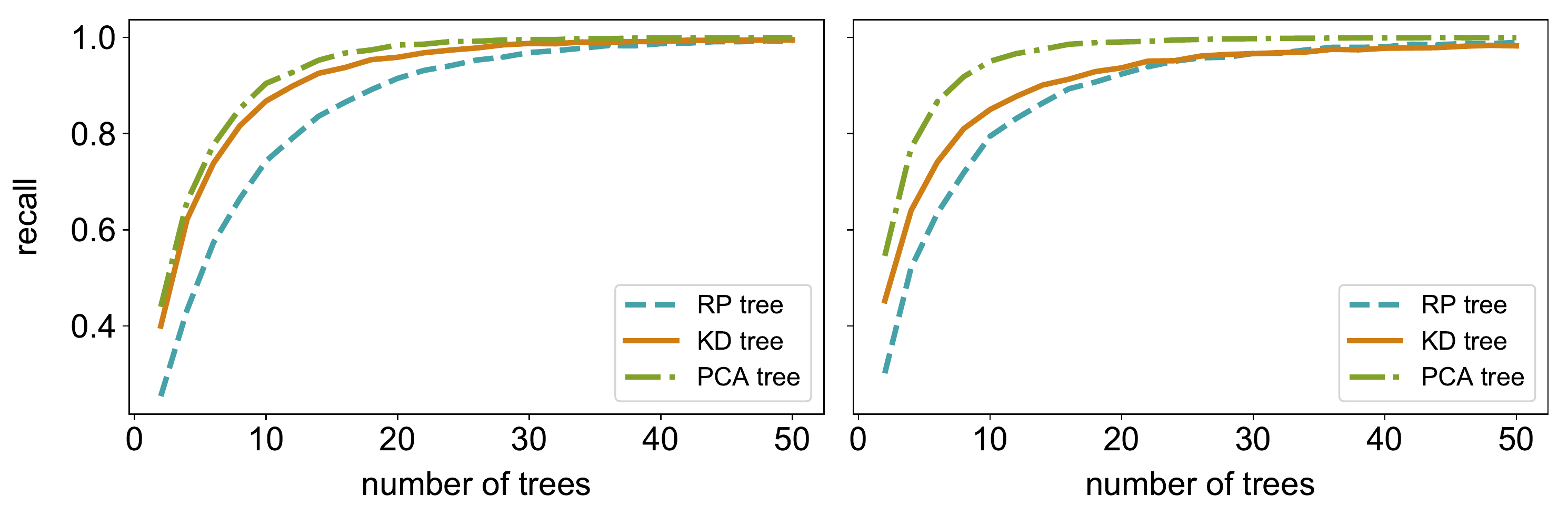}
     \caption{Recall (for $k = 10$) as a function of the number of trees on MNIST (left) and Fashion-MNIST (right) for RP, RKD and randomized PCA trees. $\ell = 8, v = 1$.}
     \label{fig:tree-types}
\end{figure}

\subsection{Comparison of randomization and search methods}
\label{sec:tree-comparison}

Figure~\ref{fig:pq_vs_voting} shows the accuracy--speed trade-off for all combinations of the considered tree types and search methods on two benchmark data sets. For RP trees, the results are in line with previous experiments~\cite{Hyvonen2016}. For each type of tree, voting outperforms priority queue (for a given recall level, its query time is faster).


For different tree types, the results vary between the data sets. Note that although both data sets for which results are shown have the same sample size ($n=60000$) and dimensionality ($d=784$), the relative order of the trees is different: for MNIST, RKD trees are the fastest, and randomized PCA trees are the slowest; whereas for Fashion-MNIST, randomized PCA trees are the fastest, and RKD trees are the slowest. This means that the relative performance of different randomization strategies depends also on the distribution of the data.

Figure \ref{fig:tree-types} further illustrates the differences between the tree types with fixed parameters: on Fashion-MNIST, PCA trees are noticeably more accurate than RKD trees and RP trees, especially for a small amount of trees. This explains the stronger performance of PCA trees with the optimal parameters; observe that the slightly stronger performance of RKD trees on MNIST is due to their faster projection times (1 vs. $\sqrt{d}$ operations per projection).

However, the differences between tree types are less pronounced than the difference between search methods. Since we can use the same projection vector on each node at the same level of an RP tree, they are the fastest to build (see Table 2). Thus, we present some of the experimental results only for them.


\section{An autotuning algorithm}
\label{sec:autotuning}

Since voting outperforms using a priority queue for all the data sets and the tree types, we present an autotuning algorithm for the voting search. Any of the different tree types can be used. Hence, the tuned hyperparameters $\bm{\alpha}$ are the number of trees $T$, the depth of the trees $\ell$, and the vote threshold $v$.

The optimal hyperparameter values are searched from the whole range
$\bm{\alpha}_\mathrm{lim} = (1,  \dots, T_\mathrm{max})\times (\ell_1,\dots, \ell_\mathrm{max}) \times (1, \dots , v_\mathrm{max})$;  setting a grid interval is not required. Here we use $v_\mathrm{max} = T_\mathrm{max}$, $\ell_\mathrm{max} = \lfloor \log_2{n} \rfloor$. Since each individual tree consumes the same amount of memory and takes an equal time to grow, $T_\mathrm{max}$ can be chosen as a limit on the building time or the memory consumption.

\begin{algorithm}
\begin{algorithmic}[1]
\raggedright
\Function{generate-index-auto}{$\bm{\alpha}_\mathrm{lim}, \mathbf{x}, \mathbf{q}, k, \bm{\psi}$}
  \State trees $\leftarrow$ \Call{grow-trees}{$\mathbf{x},\bm{\alpha}_\mathrm{lim}, \bm{\psi}$}
  \For{$i = 1, \dots, m$}
  	\State true-knn $\leftarrow$ \Call{exact-knn}{$q_i, k, \mathbf{x}$}
  	\parState{$A_i$ $\leftarrow$ \Call{count-elected}{$\bm{\alpha}_\mathrm{lim}$, $q_i$, true-knn}}
  	\parState{$B_i$ $\leftarrow$ \Call{count-elected}{$\bm{\alpha}_\mathrm{lim}$, $q_i$, $\{1, \dots, n\}$}}
  \EndFor
  \State recalls $\leftarrow \frac{1}{km}\sum_{i=1}^m A_i $
  \label{line:estimating-recall}
  \State query-times $\leftarrow$ \Call{fit-times}{$ \frac{1}{m}\sum_{i=1}^m B_i$, $\mathbf{x}$.dim}  
  \label{line:estimating-candidate-set}
  \State \Return recalls, query-times, trees
\EndFunction
\end{algorithmic}
\caption{}
\label{algo:autotuning-main}
\end{algorithm}

\subsection{Estimating recall and candidate set size}
\label{sec:estimating-recall}

The autotuning algorithm (Algorithm \ref{algo:autotuning-main}) first builds an index consisting of $T_\mathrm{max}$ trees of depth $\ell_\mathrm{max}$ (function \textproc{grow-trees}). The true neighbors of each test query $q_i$ are subsequently found by the function \textproc{exact-knn}.



For each test query, the elected points are counted by \textproc{count-elected} (Algorithm \ref{algo:autotuning-count-elected}) for two sets: the whole data set and the set of true $k$ nearest neighbors. When using an index consisting of the first $T$ trees, all the points that were elected when using an index consisting of the first $T-1$ trees are also elected for the fixed vote threshold $v$. This means that we only have to count the points which get their $v$:th vote at the $T$:th tree (line \ref{line:add-votes} of Algorithm \ref{algo:autotuning-count-elected}). Hence, we can count the numbers of elected points for all $1, \dots, T_\mathrm{max}$ number of trees with minimal overhead compared to counting them only for $T_\mathrm{max}$ trees.


\begin{algorithm}
\begin{algorithmic}[1]
\raggedright
\Function{count-elected}{$\bm{\alpha}_\mathrm{lim}$, $q$, $I$}
  \State initialize three-dimensional tensor $A$
  \For{$\ell = \ell_1, \dots, \ell_\mathrm{max}$}
    \State initialize votes as zero vector of length $n$
  \State initialize c as zero vector of length $v_\mathrm{\max}$
  \For{$T = 1, \dots, T_\mathrm{max}$}
  \State $c$ $\leftarrow$ $c$ + \Call{count-votes}{$T$, $\ell$, $q$, $I$, votes} \label{line:add-votes}
  \State write $c$ to $A$
  \EndFor
  \EndFor
  \State \Return $A$
\EndFunction
\Function{count-votes}{$T$, $\ell$, $q$, $I$, votes}
\State initialize counts as zero vector of length $v_\mathrm{\max}$
\State leaf $\leftarrow$ node containing $q$ at level $\ell$ of the $T$:th tree
\For{point in leaf}
  \If{point $\in I$}
    \State votes[point] $\leftarrow$ votes[point] + 1
    \State counts[votes[point]] $\leftarrow$ counts[votes[point]] + 1
  \EndIf
\EndFor
\State \Return counts
\EndFunction
\end{algorithmic}
\caption{}
\label{algo:autotuning-count-elected}
\end{algorithm}

The counting is done by the function \Call{count-votes}{} (Algorithm \ref{algo:autotuning-count-elected}) which adds a vote for each point of the node, and for each $v = 1, \dots v_{\mathrm{max}}$, counts how many points of this node get their $v$:th vote.

Finally, the expected recall and candidate set size can be estimated by their sample means for each parameter combination (lines \ref{line:estimating-recall} and \ref{line:estimating-candidate-set} in Algorithm \ref{algo:autotuning-main}). 
Since a brute force strategy of performing actual test queries and timing them for each possible hyperparameter combination in the set $\bm{\alpha}_\mathrm{lim}$ is impractically slow, the function \textproc{fit-times} estimates the expected query time as a function of the candidate set size and data dimension as described in the following section.

\subsection{Estimating the query time}
\label{subsec:estimating-query}

We exploit linear scaling of the components of a query to build a model which estimates the query time. The query time can be split into the candidate pruning time and the final search time. Further, the candidate pruning phase is dominated by two operations: projecting the points onto the split directions, and vote counting. This suggests that we can estimate each of the three times separately:
\[
\mathrm{Time}(q;\bm{\alpha},\mathbf{r}) \approx   \mathrm{Time}_{\mathrm{proj}}(q;\bm{\alpha},\mathbf{r})+ \mathrm{Time}_{\mathrm{vote}}(q;\bm{\alpha},\mathbf{r})+ 
\mathrm{Time}_{\mathrm{dist}}(q;\bm{\alpha},\mathbf{r}).
\]


\subsubsection{Projection time}
The projection time depends on the type of randomization used in the trees. In RKD trees, the projection time is insignificant because coordinate axes are used as split directions. For RP trees and randomized PCA trees, the query point is projected onto a sparse vector at each level of each tree. Hence, the projection time is approximately linear w.r.t.\ the number of random vectors $z := T\ell$ the query point is projected onto. Thus, we can use a linear model to estimate the projection time for known hyperparameters $T$ and $\ell$.

To collect the data for the model, we design an experiment by choosing a representative sample $\mathbf{z} = (z_1, \dots, z_w)$ of sparse random matrices with $d$ columns and $z_1, \dots, z_w$ total components, and measuring the elapsed times to multiply a $d$-component vector by each of these matrices. The sparsity is fixed as $a = 1/\sqrt{d}$.


When measuring running times, we observed that the random variation is typically small, but sometimes outliers appear, for example due to other processes activating on the background. This is why we use the Theil-Sen estimator \cite{Theil1992} to model the dependence between the number of random vectors and projection time. It is a non-parametric estimator for a linear trend, and is much more robust against outliers than ordinary least squares regression.


Now the expected projection time for the hyperparameter values $\bm{\alpha} = (T, \ell, v)$ can be estimated as $\widehat{\mathrm{Time}}_{\mathrm{proj}}(\mathbf{q}; \bm{\alpha}, \mathbf{r}) = \hat{\beta}_0  + z\hat{\beta}_1$,
where $z = T\ell$, and $\hat{\beta}_0$ and $\hat{\beta}_1$ are the intercept and the slope estimated by the Theil-Sen method.

\subsubsection{Voting time}

For one tree, counting the votes means adding a vote for each point of the leaf the query point falls into. For $T$ trees, this means that the whole voting step takes roughly $Tn_0$ operations, where
$n_0 = \lceil n / 2^\ell \rceil$
is the maximum leaf size. This means that we can model the voting time as a linear function of $y := Tn_0$, and proceed as in estimating the projection times.

\subsubsection{Final search time}

The final search in the candidate set is dominated by computing the distances to all $|S|$ points of the candidate set, which takes $|S|d$ operations; hence it is approximately linear with respect to the candidate set size $|S|$. Thus, we can proceed as before, this time measuring the time it takes to compute the distances from any $d$-dimensional query point to $|S|$ vectors of dimension $d$. After fitting the model, the final search time can be estimated as 
\[
\widehat{\mathrm{Time}}_{\mathrm{dist}}(\mathbf{q}; \bm{\alpha}, \mathbf{r}) = \hat{\alpha}_0  + |\bar{S}(\mathbf{q};\bm{\alpha}, \mathbf{r})|\hat{\alpha}_1,
\]
where $\hat{\alpha}_0$ and $\hat{\alpha}_1$ are the coefficients of the Theil-Sen estimator, and 
$|\bar{S}(\mathbf{q};\bm{\alpha}, \mathbf{r})|$
is the observed mean candidate set size for this hyperparameter combination.

\subsection{Using the autotuning index}

After the expected recall levels and the query times have been computed, finding the optimal parameter combination is a matter of a simple table lookup. Since the index has already been built, growing new trees is not required: if the optimal parameter combination is $\hat{\bm{\alpha}} = (\hat{T}, \hat{\ell}, \hat{v})$, we can just pick the first $\hat{T}$ trees that have already been built, and prune them to depth $\hat{\ell}$.

\begin{figure}[!t]
    \centering
    \includegraphics[scale=0.48,clip=yes,trim=30 10 30 10]{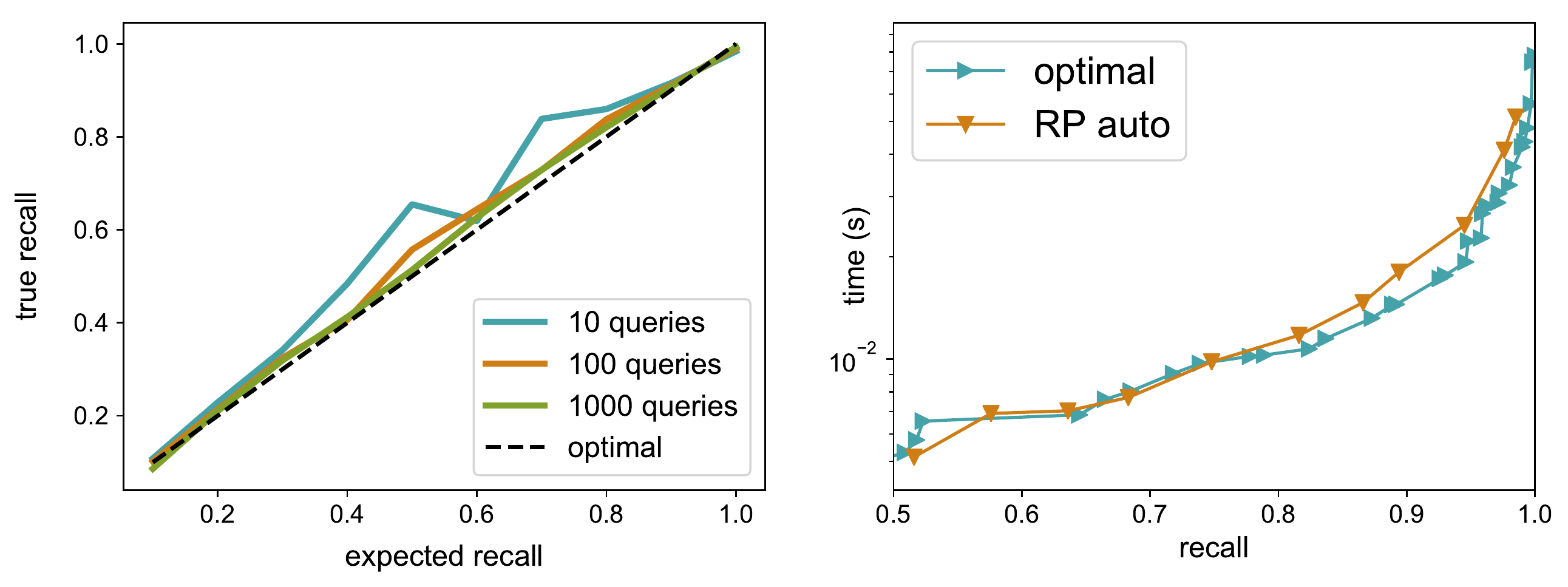}
    \caption{Left: Recall estimated by autotuning vs. recall on the test set. Right: Recall vs. query time on test set for optimal parameters and auto-tuned parameters. $k$=10.}
    \label{fig:acc}
\end{figure}

\section{Experimental results}
\label{sec:experimental-results}

First, we verify using RP trees that the autotuning algorithm accurately estimates the recall. Figure~\ref{fig:acc} (a) shows estimated recall on a validation set against recall on an independent test set for the MNIST data set. Larger validation sets yield sharper estimates, indicating the consistency of the estimator. The results are similar for other data sets and tree types. Figure~\ref{fig:acc} (b) compares on an independent test set hyperparameters optimized by the autotuning algorithm (RP auto) for the validation set to hyperparameters optimized for the test set (optimal). The parameters found by the algorithm are near-optimal. 

Next, we compare the performance of the presented algorithm with RP trees to other autotuning algorithms for ANN: autotuning for VP-trees~\cite{Yianilos1993} and multi-probe LSH~\cite{Dong2008} implemented in NMSLib~\cite{Boytsov2013} and autotuning for RKD trees and hierarchical k-means trees in FLANN~\cite{Muja2014}. To the best of our knowledge, these are the only available ANN libraries that feature an autotuning method. The compared libraries and our own code are all written in C++.
\newcolumntype{L}[1]{>{\raggedright\let\newline\\\arraybackslash\hspace{0pt}}m{#1}}
\newcolumntype{C}[1]{>{\centering\let\newline\\\arraybackslash\hspace{0pt}}m{#1}}
\newcolumntype{R}[1]{>{\raggedleft\let\newline\\\arraybackslash\hspace{0pt}}m{#1}}
\renewcommand{\arraystretch}{1.05}
\begin{table}[t]
\begin{center}
\begin{tabular}{  r  l C{1.1cm} C{1cm} C{1.15cm} C{1.15cm} | C{1.1cm} C{1cm} C{1.15cm} C{1.15cm} }
& & \multicolumn{4}{c|}{Target recall 80\%} & \multicolumn{4}{c}{Target recall 90\%} \\
\cline{3-10}
& & RP & LSH & VPtree & FLANN & RP & LSH & VPtree & FLANN \\
\hline

\multirow{3}{4em}{MNIST} & tuning & \bf{13.23} & 26.84 & 744.4 & 102.2 & \bf{13.23} & 24.61 & 926.1 & 113.9 \\ 
& search & \bf{0.111} & 1.164 & 0.739 & 0.206 & \bf{0.169} & 2.513 & 1.368 & 0.311 \\ 
& recall & \textbf{0.822} & 0.853 & 0.831 & 0.654 & \textbf{0.909} & 0.939 & 0.911 & 0.790 \\
& stdev & $\pm$0.009 & -- & -- & $\pm$0.020 & $\pm$0.004 & -- & -- & $\pm$0.017 \\  
\hline

\multirow{3}{4em}{Fashion} & tuning & \bf{13.22} & 26.70 & 396.5 & 104.8 & \bf{13.23} & 25.38 & 427.4 & 136.4 \\ 
& search & \bf{0.129} & 0.917 & 0.353 & 0.310 & \bf{0.198} & 1.575 & 0.557 & 0.216  \\ 
& recall & \textbf{0.798} & 0.850 & 0.813 & 0.693 & 0.881 & 0.927 & \textbf{0.908} & 0.825 \\ 
& stdev & $\pm$0.007 & -- & -- & $\pm$0.034 & $\pm$0.006 & -- & -- & $\pm$0.025 \\  
\hline

\multirow{3}{4em}{Trevi} & tuning & \bf{75.89} & 156.1 & 3026 & 724.9 & \bf{76.28} & 158.9 & * & 751.6 \\ 
& search & \bf{1.730} & 14.01 & 13.58 & 2.813 & \bf{3.371} & 25.63 & * & 4.276  \\ 
& recall & \textbf{0.822} & 0.837 & 0.832 & 0.566 & \bf{0.914} & 0.918 & * & 0.679 \\ 
& stdev & $\pm$0.011 & -- & -- & $\pm$0.028 & $\pm$0.006 & -- & * & $\pm$0.016 \\  
\hline

\multirow{3}{4em}{Random} & tuning & \bf{32.78} & 55.48 & 120.6 & 134.6 & \bf{32.76} & 54.56 & 134.0 & 149.1 \\ 
& search & 0.074 & 0.256 & 0.409 & \textbf{0.049} & 0.095 & 0.249 & 0.659 & \textbf{0.087} \\ 
& recall & \textbf{0.804} & 0.882 & 0.827 & 0.602 & \textbf{0.902} & 0.941 & 0.911 & 0.728 \\ 
& stdev & $\pm$0.012 & -- & -- & $\pm$0.015 & $\pm$0.007 & -- & -- & $\pm$0.015 \\  
\hline
\multirow{3}{4em}{GIST} & tuning & \bf{317.4} & 484.1 & 960.4 & * & \bf{318.1} &  437.9 & 1127 & * \\ 
& search & \bf{9.253} & 122.4 & 41.55 & * & \bf{15.51} & 205.7 & 66.54 & * \\ 
& recall & \textbf{0.784} & 0.862 & 0.864 & * & \textbf{0.881} & 0.942 & 0.940 & * \\ 
& stdev & $\pm$0.011 & -- & -- & * & $\pm$0.005 & -- & -- & * \\  
\hline
\end{tabular}
\end{center}
\caption{Comparison of autotuning algorithms. Autotuning times (seconds), query times for 1000 queries (s) and recall (for $k=10$) measured on a test set (* = did not complete within one hour). For the randomized algorithms (RP and FLANN), average recalls of 10 runs with the corresponding standard deviations are reported. The best result in each case is typeset in boldface.} 
\end{table}
\begin{figure}
    \centering
    \includegraphics[scale=0.43,clip=yes,trim=50 10 30 10]{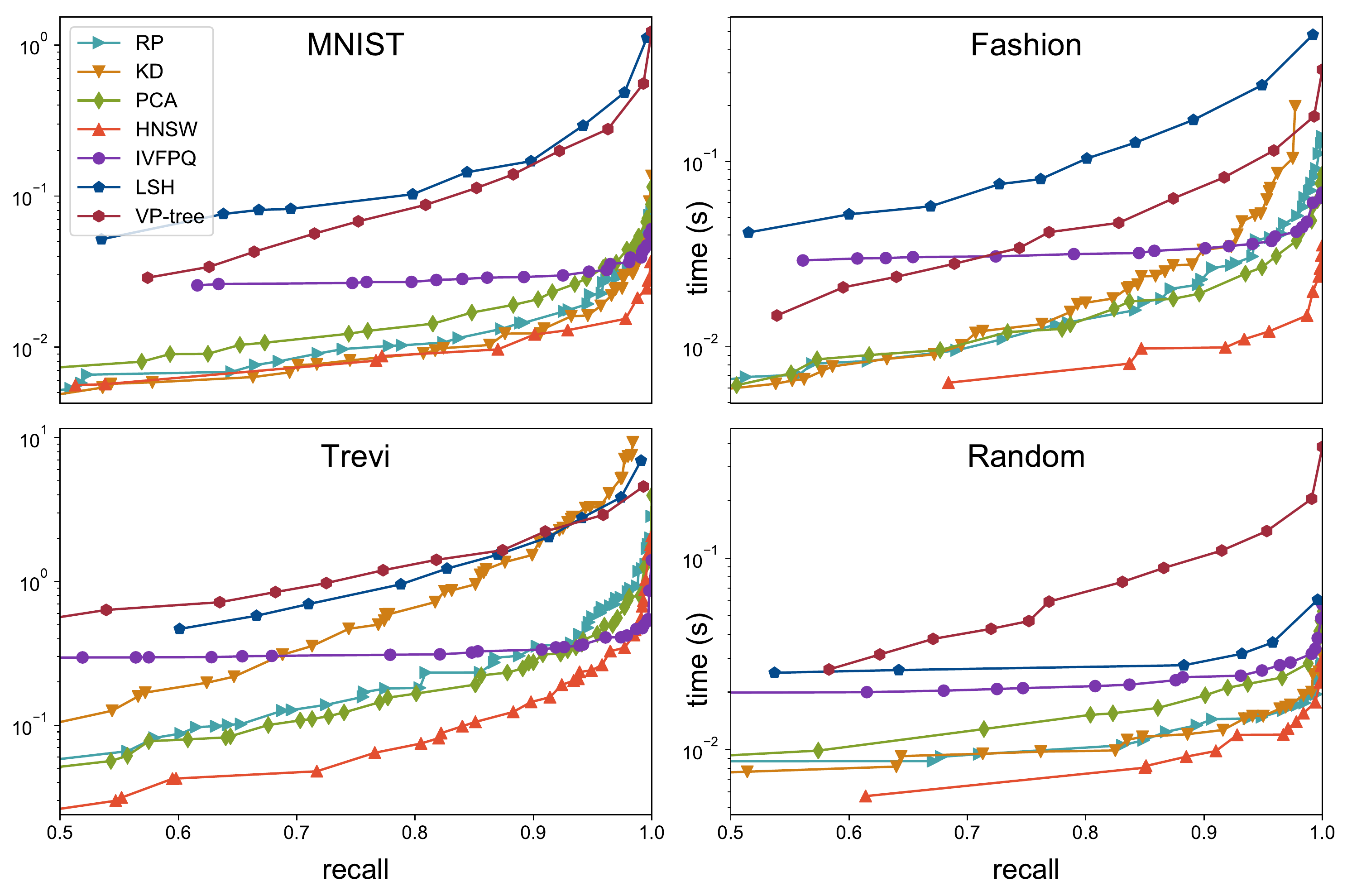}
    \caption{Recall vs. query time (s) for 100 queries for different ANN algorithms. $k=10$.}
    \label{fig:comparison}
\end{figure}

The data sets used in the experiments are MNIST ($n$=60000, $d$=784), Fashion-MNIST ($n$=60000, $d$=784), Trevi ($n$=101120, $d$=4096), Random ($n$=256000, $d$=256) and GIST ($n$=1000000, $d$=960). Table 1 shows for two target recall rates (0.8 and 0.9) the autotuning time (including the index-building time), and the query time and recall on a test set which was not used to tune the hyperparameters. The proposed tuning algorithm (with RP trees) is fastest at index building in all cases. Our approach has significantly faster query times than VP trees and LSH in all cases, and faster query times than FLANN for all but one data set. Our approach is also the most accurate at estimating the recall in most cases. The other methods systematically over- or underestimate the recall.



We also compare tree-based ANN against state-of-the-art quantization-, and graph-based algorithms: IVFPQ~\cite{Jegou2011} and HNSW~\cite{Malkov2016} implemented in FAISS \cite{Johnson2017}. As autotuning for these methods is not available, we perform a grid search on the possible parameter values. Figure~\ref{fig:comparison} shows that tree-based methods are faster than PQ (except on highest recall levels) and close to the performance of HNSW. We emphasize that according to an independent benchmarking project\footnote{\url{https://github.com/erikbern/ann-benchmarks}}, HNSW is the fastest ANN algorithm available. Multi-probe LSH and VP tree are also included in the comparison; they are significantly slower than the other methods.%

Finally, we compare the index building time (Table 2). Even though HNSW has faster query times, RP trees are significantly faster to build. The whole autotuning takes less time than building a single HNSW index. We emphasize that these results are on a single thread; the differences become more pronounced with multiple threads as the indexing process is embarrassingly parallel for trees. An implementation of the proposed algorithm is available in the MRPT library\footnote{\url{https://github.com/vioshyvo/mrpt}}.

\subsubsection*{Acknowledgments.} This project was supported by Business Finland (project 3662/31/2018 Advanced Machine Learning for Industrial Applications) and the Academy of Finland (project 311277 TensorML).

\begin{center}
\begin{table}[!t]
\label{table:index-building-time}
\centering
\begin{tabular}{ l C{1.5cm} C{1.5cm} C{1.5cm} C{1.5cm} C{1.5cm}}
        & RP      & RKD      & PCA   & HNSW  & IVFPQ \\
        \hline
MNIST   & \textbf{3.62}    & 7.40     & 12.63 & 25.1 & 27.31 \\
Fashion & \textbf{1.86}    & 14.3    & 13.12 & 20.2 & 30.71 \\
Trevi   & 102   & \textbf{43.2}    & 185 & 266 & 262.5 \\
Random  & \textbf{2.23}    & 6.83     & 27.8 & 90.3 & 63.59 \\
\end{tabular}
\caption{Index building times (seconds) for optimal parameters at 90\% recall for different ANN algorithms. The best result on each data set is typeset in boldface.}
\end{table}
\end{center}











%
%
%

\bibliographystyle{splncs04}
\bibliography{RP_trees_theory_refs}

\end{document}